\documentclass[aps,superscriptaddress,amsmath,showpacs,twocolumn,pra]{revtex4}
\usepackage[T1]{fontenc}
\usepackage[latin9]{inputenc}
\usepackage{amssymb}
\usepackage{graphicx}
\usepackage{epsfig}
\usepackage{txfonts}
\usepackage{amsfonts}
\usepackage{dcolumn}
\usepackage{esint}
\usepackage{mathrsfs}
\usepackage{bm}

\begin{document}

\title{Stopping waveguide photons with controllable atomic cavity}

\author{Xingmin Li}

\affiliation{State Key Laboratory of Optoelectronic Materials and Technologies,
School of Physics, Sun Yat-Sen University, Guangzhou
510275, China}

\author{L. F. Wei\footnote{E-mail: weilianfu@gmail.com}}

\affiliation{State Key Laboratory of Optoelectronic Materials and Technologies,
School of Physics, Sun Yat-Sen University, Guangzhou
510275, China}

\affiliation{Quantum Optoelectronics Laboratory, School of Physical Science and
Technology, Southwest Jiaotong University, Chengdu 610031, China}

\begin{abstract}
Slowing/stopping the light travelling in free space with electromagnetically induced transparency to implement the optical quantum information processings and store information has been paid much attention in recent years. For the waveguide photons, here, we propose an approach to stop them with a controllable atomic cavity generated by a pair of atomic mirrors; one of them reflects the photon completely and another one with the adjustable reflected/transmitted probability of the photons. Based on the full quantum mechanical theory in real space, we show that the reflected/transmitted probability of the waveguide photon with a fixed frequency can be controlled by adjusting the energy-splitting of the driven two-level atomic scatters (i.e., atomic mirrors). As a consequence, the photon can be controllably transmitted/reflected along the waveguide by the aside atomic mirrors with the adjustable atomic energy levels. Ideally, the photons could be stopped in the atomic cavity. This provides a novel mechanism to stop/retravel the waveguide photon in a controllable ways. The feasibility of the proposal with the current integrated optical devices is also discussed.
\end{abstract}

\maketitle


Single photons propagating is a basic and important subject in quantum optics, and associates with many quantum information and quantum computation processes~\cite{Biolatti PRL 2000,Lukin PRL 2001,Lopez PRL 2008}. And many theoretical and experimental works are proposed to investigate the processes of the single photons transporting in one-dimension waveguide and coupling to two- and multi-level quantum systems~\cite{Wallraff Nature 2004,Shen PRL 2005,Shen PRA 2009,Zheng PRA 2010,Zheng PRL 2013}. Following the investigations of single photons propagating, some works on quantum-detection, quantum-routing and quantum-switch of single photons in one-dimension waveguide are proposed in theoretical and even realized in experimental~\cite{Romero PRL 2009,Chen PRL 2011,Li PRA1 2015,Zhou PRL 2013,Li PRA2 2015,Liao PRA 2009}. However, in these schemes, the two- and multi-level quantum systems all have the time-independent energy frequencies, which means we can not adjust the detuning between the single photons and the quantum systems when they are integrated in a chip. Meanwhile, the transmitted and reflected probabilities of the single photons also can not be adjusted as convenient as we can. How to perfectly solve this work is a challenge, and the time-dependent quantum system must be needed.

In the past few years, the time-dependent quantum dynamics in low-dimensional systems are always attraction. Many methods are proposed in theoretical to investigate the time-dependent external fields controlling the quantum system~\cite{Torres APL 2009,Chen PRA 2015,Yan PhysB 2012}. However, these investigations are mostly focus on the transporting of electrons~\cite{Yan PhysB 2012,Gudmundsson PRB 2012} and very little on photons~\cite{Yu PRA 2014,Mirza PRA 2014}. Therefore, single photons propagating in one-dimensional waveguide and coupling to a time-dependent quantum system deserves more exploration. The investigations on transporting of electrons allow us to directly construct the similar quantum scheme with energy variable atoms (special artificial atoms). In order to distinctly study transmission relationship in time-dependent quantum system, we propose a full quantum-mechanical approach to solve that single photons couple to energy variable system (i.e. artificial atom controlled by voltage or flux) depending on time in one-dimension waveguide. Because of the time-dependent quantum system, the single photons will be separated into different sidebands and the total transmitted probability of the single photons is the sum of the transmitted probability in each sideband. The total transmitted probability is seriously depending on the variable energy's parameters, such as amplitude and frequency of the vibration. Also this character lets the control of the single photons transporting become more conveniently.

We present our model of a single photon coupling to the energy variable system in one-dimension waveguide, and calculate the transmitted and reflected amplitudes in different sidebands.
 And show the analysis of the transmission spectra for different parameters and compare transmitted probabilities of the zeroth and first sideband. Finally, a feasible design is given 


We consider a single-photon transporting in an one-dimension waveguide, and being scattered by a time-dependent aside ideal two-level quantum system (i.e., two-level artificial atom). We sketch the entire scheme in Fig.~\ref{fig:1}, and set the atom located at $x_0$. Then the Hamiltonian can be written as ($\hbar=1$) :

\begin{figure}[b]
\centering{}
\includegraphics[width=0.36\textwidth]{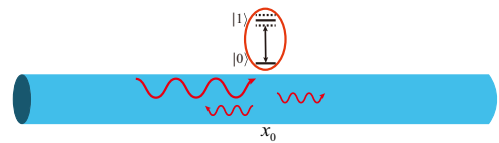}
\caption{\label{fig:1} A single-photon transport in an one-dimension waveguide, and a time-dependent two-level quantum system is located at $x_{0}$.}
\end{figure}

\begin{eqnarray}
H & = & \int dx\left[c_{R}^{\dagger}(x)(-iv_{g}\frac{\partial}{\partial x})c_{R}(x)+c_{L}^{\dagger}(x)(iv_{g}\frac{\partial}{\partial x})c_{L}(x)\right]\nonumber \\
 &  & +\int dxV\delta(x-x_{0})\left[c_{R}(x)\sigma^{+}+c_{L}(x)\sigma^{+}+H.c.\right]\nonumber\\
 &  & +\Omega\left[1+f\cos(\omega t)\right]\sigma^{+}\sigma^{-}.\label{eq:1}
\end{eqnarray}
Here, $c_{R}^{\dagger}(x)$ ($c_{R}(x)$) and $c_{L}^{\dagger}(x)$ ($c_{L}(x)$) are the bosonic
creation (annihilation) operators of the single-photon pulse propagating right and left, respectively. $v_{g}$ is the group velocity of the photon, $V$ is the coupling strength between the waveguide photon
and the atom, and $\Omega$ the atomic transition frequency between the ground and excited states (with $\sigma^{+} (\sigma^{-})$ being the atomic raising (lowering) ladder operator). $\omega$ and $f$ are the variation frequency and amplitude of the two-level atom. In order to keep the coupling strength $V$ unchangeable, the amplitude satisfies $f\ll 1$. Because of considering an ideal quantum system, we neglect the dissipations of the atom and the waveguide.

The most general wave function can be expressed as
\begin{eqnarray}
|\Psi(x,t)\rangle
& = & \int dx\left[\phi_{R}(x,t)c_{R}^{\dagger}(x)+\phi_{L}(x,t)c_{L}^{\dagger}(x)\right]|\varnothing \rangle\nonumber \\
 &  & +e(t)\sigma^{+}|\varnothing \rangle, \label{eq:2}
\end{eqnarray}
with $|\varnothing \rangle $ being the ground state, i.e., without any photon in the waveguide and the atom
stays at its ground state $|0\rangle $. $\phi_{R/L}(x,t)$ and $e(t)$ stand for the time-dependent probabilistic amplitudes of the photon propagating along the $R/L$ direction and the atomic excitation, respectively.

Substitute the functions Eqs.~\eqref{eq:1} and ~\eqref{eq:2} into the time-dependent
Schr{\"o}dinger equation
\begin{equation}
i\partial_{t}|\Psi(x,t)\rangle =H|\Psi(x,t)\rangle,\label{eq:3}
\end{equation}
and solve the equation. Corresponding the coefficient of each element, we can get the following set of equations:

\begin{subequations}
\begin{align}
& i\frac{\partial}{\partial t} \phi_{R}(x,t)  = -iv_{g}\frac{\partial}{\partial x}\phi_{R}(x,t)+V\delta(x)e(t),\label{eq:3a}\\
& i\frac{\partial}{\partial t} \phi_{L}(x,t) = iv_{g}\frac{\partial}{\partial x}\phi_{L}(x,t)+V\delta(x)e(t).\label{eq:3b}\\
& i\frac{\partial}{\partial t} e(t) = \Omega\left[1+f\cos(\omega t)\right]e(t)+V\left[\phi_{R}(0,t)+\phi_{L}(0,t)\right].\label{eq:3c}
\end{align}
\end{subequations}

 While the input single photon is scattered by the time-dependent atom, some sidebands will be appeared. And the single photon  can be reflected and transmitted though these sidebands with certain probabilities. The above probabilistic amplitudes of the photon propagating along the $R/L$ direction can be further expressed as
\begin{subequations}
\begin{align}
& \phi_{R}(x,t)  =  \theta(-x+x_{0})e^{i(q_{0}x-\omega_{0}t)}+\theta(x-x_{0})\sum_{n}e^{i(q_{n}x-\omega_{n}t)}t_{n},\label{eq:4a}\\
& \phi_{L}(x,t)  =  \theta(-x+x_{0})\sum_{n}e^{-i(q_{n}x+\omega_{n}t)}r_{n},\label{eq:4b}
\end{align}
\end{subequations}
where $t_{n}$ and $r_{n}$ stand for the transmitted and reflected
amplitudes of the photon in the $n-$th sideband, respectively. $\omega_{0}$ ($\omega_{n}$) is the frequency (in $n-$th sideband) of the input single photon, and $q_{0}$ ($q_{n}$) is the wave vector (in $n-$th sideband).

Without loss of the generality, we take $x_{0}=0$ for simplicity. Then, by substituting Eq.~\eqref{eq:4a}-~\eqref{eq:4b} into Eq.~\eqref{eq:3a}-~\eqref{eq:3c}, we have the following equations:
\begin{subequations}
\begin{align}
& \sum_{n}e^{-i\omega_{n}t}t_{n} =  \sum_{n}e^{-i\omega_{n}t}r_{n}+e^{-i\omega_{0}t}, \label{eq:5a}\\
& Ve(t) = iv_{g}\sum_{n}e^{-i\omega_{n}t}r_{n},\label{eq:5b}\\
& i\frac{\partial}{\partial t} e(t) = \Omega\left[1+f\cos(\omega t)\right]e(t)+V\left[\sum_{n}e^{-i\omega_{n}t}r_{n}+e^{-i\omega_{0}t}\right].\label{eq:5c}
\end{align}
\end{subequations}
with $q_{n}=\omega_{n}/v_{g}$. Using the Jacobi-Anger expansion of Bessel function~\cite{Abramowitz}:
\begin{eqnarray}
e^{iu\sin x}=\sum_n J_{n}(u)e^{inx}, \label{eq:6}
\end{eqnarray}
with $J_{n}(u)$ being the first kind Bessel function of order $n$. We can express the transmitted and reflected amplitudes as:
\begin{subequations}
\begin{align}
& r_{n}=\sum_{l}\frac{-i\gamma J_{l}(\frac{f\Omega}{\omega})J_{n+l}(\frac{f\Omega}{\omega})}{\Delta-l\omega+i\gamma},\label{eq:7a}\\
& t_{n}=\sum_{l}\frac{-i\gamma J_{l}(\frac{f\Omega}{\omega})J_{n+l}(\frac{f\Omega}{\omega})}{\Delta-l\omega+i\gamma}+\delta_{n,0},\label{eq:7b}\\
& e(t) = \sum_{n,l} \frac{-i\gamma e^{-i\omega_{n}t}}{\Delta-l\omega+i\gamma}J_{l}(\frac{f\Omega}{\omega})J_{n+l}(\frac{f\Omega}{\omega}),\label{eq:7c}
\end{align}
\end{subequations}
with $\Delta=\omega_{0}-\Omega$, $\gamma={V^2}/{v_g}$ and $\omega_{n}=\omega_{0}+n\omega$. Here, $\Delta$ and $\gamma$ are the detuning and the effective coupling strength between the photon and time-dependent atom, respectively. Note that $\gamma$, $\Delta$ and $\omega$ all have the same unit as frequency, and effective coupling strength satisfies $\gamma\ll \Omega$.

Because of the periodical variation of the energy of the atom, the transmitted and reflected amplitudes are separated into several parts. Thus the total transmitted and reflected probabilities should be written as:
\begin{subequations}
\begin{align}
& R =\sum_{n}|t_{n}|^2,\label{eq:8a}\\
& T =\sum_{n}|r_{n}|^2.\label{eq:8b}
\end{align}
\end{subequations}
In the time-dependent quantum system, the transmitted and reflected probabilities also satisfy: $T+R=1$. In the following section, we discuss the transmission spectra of the single photon scattered by the time-dependent atom.


Here we will analyse the transmitted spectra of the single photon by two cases: $f\omega=0$ and $f\omega \neq 0$.

\begin{figure}[t]
\centering{}
\includegraphics[width=0.36\textwidth]{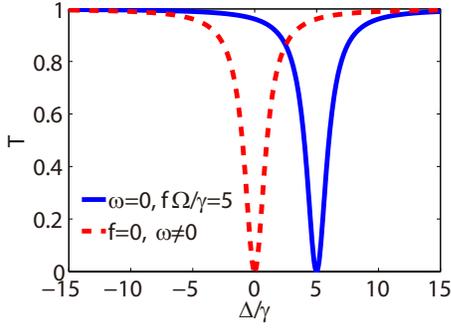}
\caption{\label{fig:2} The time-dependent atom with a very low frequency or very trivial amplitude. For simplicity is $f\omega=0$.}
\end{figure}
\begin{figure}[t]
\centering{}
\includegraphics[width=0.48\textwidth]{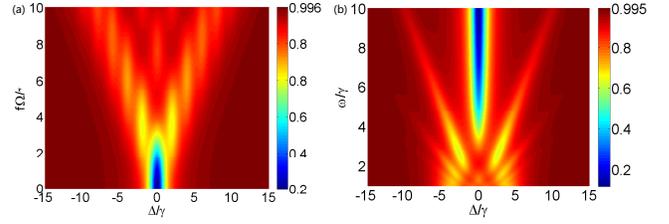}
\caption{\label{fig:3} The transmission spectra of the single photon coupling with the time-dependent atom. (a) Take the amplitude $f$ as the variable parameter and fix the frequency in $\omega/\gamma=2$. (b) Take the frequency $\omega$ as the variable parameter and fix the amplitude in $f\Omega/\gamma=5$.}
\end{figure}

i) $f\omega=0$. In this case, we can consider that the time-dependent atom with very low frequency (i.e., $\omega \sim 0$) or very trivial amplitude (i.e., $f\sim 0$). Then the transmission spectra are shown in Fig.~\ref{fig:2}. The blue solid line expresses that the time-dependent atom has low frequency $\omega=0$ and un-neglected the amplitude $f\Omega/\gamma=5$. Then we can find that the complete reflected point corresponding the detuning $\Delta/\gamma=5$. In this case, the amplitude and the detuning satisfies $f\Omega=\Delta$, which also obey the principle that the resonant input single photon is complete reflected. The red dashed line show that the time-dependent atom has trivial amplitude ($f=0$) and nonzero frequency ($\omega\neq0$), and it has no different with coupling to a normal two-level atom with transmission frequency $\Omega$.

ii) $f\omega \neq 0$. When the single photon is scattered by the time-dependent atom, the transmission spectra have obvious difference with the time-independent one. The transmission spectra are shown in Fig.~\ref{fig:3}. In the Fig.~\ref{fig:3} (a), we keep the frequency unchanged (i.e.,$\omega/\gamma=2$), and adjust the amplitude $f$. For the resonant input photon, the complete reflected phenomenon is only corresponding the low amplitude (i.e., $0<f\Omega/\gamma <2$), then the transmitted probability is increasing with the increasing of the amplitude even reach complete transmission. For the non-resonant input photon, there are also some summits of reflection corresponding the relatively large detuning, and that is different form that nearly complete transmission when a single photon couple to a time-independent atom. In the Fig.~\ref{fig:3} (b), we keep the amplitude fixing in $f\Omega/\gamma=5$, and adjust the frequency $\omega$. For the resonant input photon, the complete reflected phenomenon is corresponding the relatively high frequency (i.e., $\omega/\gamma>6$), and the transmitted probability is very large for the low frequency. For the non-resonant input photon, there are also some obvious splits corresponding the low frequency. Therefore, we can get that the anomalistic transmission spectra correspond the large amplitude and low frequency of the time-dependent atom. Note that the maximum transmitted probabilities in Fig.~\ref{fig:3} (a) and (b) are not $1$, and that is limited by the values of emulation parameters.

\begin{figure}[t]
\centering{}
\includegraphics[width=0.48\textwidth]{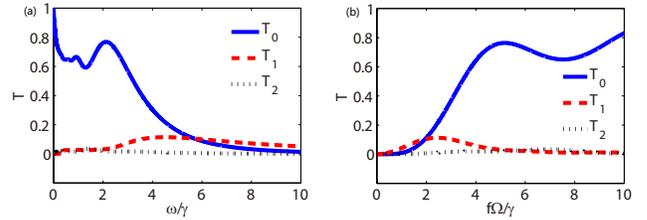}
\caption{\label{fig:4} The transmitted probabilities in different sidebands for the resonant input photons ($\Delta=0$). (a) The amplitude is fixed in $f\Omega/\gamma=5$, and (b) the frequency is fixed in $\omega/\gamma=2$.}
\end{figure}

In order to further understand the scattering process for the resonant input photons coupling to the time-dependent atom, we investigate the transmitted probabilities in different sidebands as shown in Fig.~\ref{fig:4}. When a single resonant photon is scattered by a time-independent atom, it is complete reflected. For the time-dependent atom, it is related to the sideband scattering. We fix the amplitude in $f\Omega/\gamma=5$, and the transmission curves of the different frequencies in different sidebands are shown in Fig.~\ref{fig:4} (a). The blue solid line describes the transmitted probability $T_0$ in zeroth sideband. We can find that the transmission mainly depends on the zero sidebands for the low frequency and it is decreasing with the increasing of the frequency. The red dashed line and the black dotted line show the transmitted probabilities $T_1$ and $T_2$ in the first and second sideband, respectively. We also can see that the first sideband contribute more transmitted probability than the zeroth sideband for the high frequency (i.e., $\omega/\gamma>6$) though the second sideband has trivial contribution. In the Fig.~\ref{fig:4} (b), we show the transmitted probabilities with the amplitude parameter $f$ while fix the frequency in $\omega/\gamma=2$. Here, the transmitted probability $T_0$ (blue solid line) in zeroth sideband is mainly increased with the increasing of the amplitude $f$, and becomes the main part of total transmitted probability at the large amplitude area (i.e., $f\Omega>4$). The transmitted probability in the first sideband $T_1$ (red dashed line) is larger than that in the zeroth sideband corresponding the low amplitude (i.e., $f\Omega/\gamma<2$). And there also is trivial transmitted probability in second sideband $T_2$ (black dotted line), therefore, the transmitted probabilities in high level sideband can even be neglected.


In this paper, we investigate the single photons scattered by the time-dependent quantum system in one-dimension waveguide. We find that the single photon coupling to time-dependent two-level atom can product a lot of sidebands, and these sidebands take the total transmitted probability different from the single photons transporting in one-dimension waveguide coupling with time-independent two-level atom. The complete reflection not always appeared at the resonant input single photons, and even there can appear large transmitted probabilities. We also investigate the contribution of transmission in different sidebands, and we find the zeroth and the first sidebands contribute much more transmitted probabilities than the other high level sidebands. Meanwhile, we show that the first sideband sometime play a more important role than the zeroth sidebands in transmission when we care the high frequency and low amplitude.

\begin{figure}[t]
\centering{}
\includegraphics[width=0.36\textwidth]{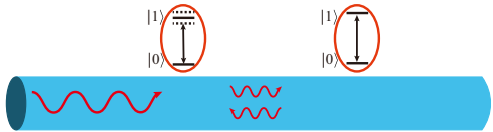}
\caption{\label{fig:5} Trap a single photon by time-dependent atom and time-independent atom.}
\end{figure}

Immediately, the approach can demonstrate its utility in optical quantum process, e.g., trap single photon see Fig.~\ref{fig:5}. Here, we propose that time-dependent and time-independent two-level atoms are located in the left and right, respectively. When cancel the external control field, we set the both quantum systems are resonant with the input single photons. Let the amplitude of the external field is large enough at first, so the input single photons can be complete transmitted. When the input photons transport the time-dependent (the left one) atom, the external control field is canceled. Because the single photons can be complete reflected by the resonant two-level atom, the single photons will be trapped in the both two-level atoms. If we want the single photon continue to propagate, another external control field can be applied on the right atom. Also, we hope more applications can be proposed based on single photons propagating in one-demission waveguide coupling to time-dependent two-level atom.

{\bf Acknowledgements.}
This work is supported by National Science Foundation of China under Grants No. 11174373, No. 91321104, and No. U1330201.


\begin{thebibliography}{10}

\bibitem{Biolatti PRL 2000} E. Biolatti, R. C. Iotti, P. Zanardi, and F. Rossi, Phys. Rev. Lett. {\bf 85}, 5647 (2000).

\bibitem{Lukin PRL 2001} M. D. Lukin, M. Fleischhauer, R. Cote, L. M. Duan, D. Jaksch, J. I. Cirac, and P. Zoller, Phys. Rev. Lett. {\bf 87}, 037901 (2001).

\bibitem{Lopez PRL 2008} L. Lopez, N. Treps, B. Chalopin, C. Fabre, and A. Ma{\^i}tre, Phys. Rev. Lett. {\bf 100}, 013604 (2008).

\bibitem{Wallraff Nature 2004} A. Wallraff, D. I. Schuster, A. Blais, L. Frunzio, R.- S. Huang, J. Majer, S. Kumar, S. M. Girvin, and R. J. Schoelkopf, Nature {\bf431}, 162-167 (2004).

\bibitem{Shen PRL 2005} Jung-Tsung Shen and Shanhui Fan, Phys. Rev. Lett. {\bf95}, 213001 (2005).

\bibitem{Shen PRA 2009} Jung-Tsung Shen and Shanhui Fan, Phys. Rev. A {\bf79}, 023837 (2009).

\bibitem{Zheng PRA 2010} Huaixiu Zheng, Daniel J. Gauthier, and Harold U. Baranger, Phys. Rev. A {\bf82}, 063816 (2010).

\bibitem{Zheng PRL 2013} Huaixiu Zheng and Harold U. Baranger, Phys. Rev. Lett. {\bf110}, 113601 (2013).

\bibitem{Romero PRL 2009} G. Romero, J. J. Garc{\'i}a-Ripoll, and E. Solano, Phys. Rev. Lett. {\bf102}, 173602 (2009).

\bibitem{Chen PRL 2011} Y.-F. Chen, D. Hover, S. Sendelbach, L. Maurer, S. T. Merkel, E. J. Pritchett, F. K. Wilhelm, and R. McDermott, Phys. Rev. Lett. {\bf107}, 217401 (2011).

\bibitem{Li PRA1 2015} Xingmin Li, Lingyun Xie, and L. F. Wei, Phys. Rev. A {\bf92}, 063840 (2015).

\bibitem{Zhou PRL 2013} Lan Zhou, Li-Ping Yang, Yong Li, and C. P. Sun, Phys. Rev. Lett. {\bf111}, 103604 (2013).

\bibitem{Li PRA2 2015} Xingmin Li and L. F. Wei, Phys. Rev. A 92, 063836 (2015).

\bibitem{Liao PRA 2009} Jie-Qiao Liao, Jin-Feng Huang, Yu-xi Liu, Le-Man Kuang, and C. P. Sun, Phys. Rev. A 80, 014301 (2009).

\bibitem{Torres APL 2009} Luis E. F. Foa Torres, and Gianaurelio Cuniberti, Appl. Phys. Lett. {\bf94}, 222103 (2009).

\bibitem{Chen PRA 2015} Chong Chen, Jun-Hong An, Hong-Gang Luo, C. P. Sun, and C. H. Oh, Phys. Rev. A {\bf91}, 052122 (2015).

\bibitem{Yan PhysB 2012} Cong Hua Yan, and Lian Fu Wei, Physica B {\bf407}, 4545šC4549 (2012).

\bibitem{Gudmundsson PRB 2012} Vidar Gudmundsson, Olafur Jonasson, Chi-Shung Tang, Hsi-Sheng Goan, and Andrei Manolescu, Phys. Rev. B {\bf85}, 075306 (2012).

\bibitem{Yu PRA 2014} Lixian Yu, Jingtao Fan, Shiqun Zhu, Gang Chen, Suotang Jia, and Franco Nori, Phys. Rev. A {\bf89}, 023838 (2014).

\bibitem{Mirza PRA 2014} Imran M. Mirza and S. J. van Enk, Phys. Rev. A {\bf90}, 043831 (2014).

\bibitem{Abramowitz} M. Abramowitz, I. A. Stegun(Eds.),Handbook of Mathematical Functions, National Bureau of Standards, Washington, D. C. , 1972.

\end{thebibliography}
\end{document}